\documentclass[11pt,fleqn]{article}

\usepackage{amsmath,amssymb,amsthm,cite,enumerate}%

\begin{document}

\begin{center}{\LARGE\bf
Quantum integrable systems \\ in three-dimensional magnetic fields: \\[0.7ex] the
Cartesian case}

\vspace{4mm} {\large Alexander Zhalij}

\vspace{2mm} { Institute of Mathematics of NAS of Ukraine, \\
3 Tereshchenkivska Str., 01601 Kyiv-4, Ukraine}

\vspace{2mm} { zhaliy@imath.kiev.ua }
\end{center}
{\vspace{5mm}\par\noindent\hspace*{8mm}\parbox{146mm}{\small
In this paper we construct integrable three-dimensional quantum-mechanical
systems with magnetic fields, admitting pairs of
commuting second-order integrals of motion. The case of Cartesian
coordinates is considered. Most of the systems obtained are new and
not related to the separation of variables in the corresponding
Schr\"odinger equation.
}\par\vspace{2mm}}

\vspace{5mm} \noindent Let us consider the stationary  Schr\"odinger
equation for particle moving in external elec\-tro\-mag\-netic field
in three-dimensional Euclidean space
\begin{equation}
H\psi=E\psi, \quad H=\dfrac 12 \vec p\, ^2+V(x,y,z)+A_i(x,y,z) p_i+
p_i A_i(x,y,z), \label{1}
\end{equation}
where $V(x,y,z)$ and $\vec A=(A_1(x,y,z), A_2(x,y,z), A_3(x,y,z))$ are
scalar and vector potentials of  electromagnetic field,
respectively. Here and below we use the notation  $\vec
p=-ih\vec\nabla$ and the summation from 1 to 3 over the repeated
indices is understood.

By analogy with classical Hamiltonian mechanics this system is
called integrable if there exists a pair of quantum-mechanical
operators $P$ and $Q$ which commute with each other as well as with
the Hamiltonian~$H$, i.e., the following relations hold
\[ [H,Q]=[H,P]=[P,Q]=0. \]
Moreover, all three operators $H,$ $Q$ and $P$ are algebraically
independent, i.e., any of them cannot be represented as polynomial
of two others~\cite{hiet1}.

In this paper we restrict ourselves by the case of  $Q$ and $P$
being quadratic polynomials of~$\vec p$
\begin{gather}\label{2}
\begin{array}{l}\arraycolsep=0ex
 Q=\alpha_{ik}(x,y,z) p_ip_k+ f_i(x,y,z) p_i+\gamma_1(x,y,z),\\[1ex]
 P=\beta_{ik}(x,y,z) p_ip_k+ g_i(x,y,z) p_i+\gamma_2(x,y,z),
\end{array}
\end{gather}

In classical mechanics integrable systems are interesting since
their motion in a phase space is more ordered, namely, it is
restricted to a torus. In quantum mechanics integrability of an
$n$-dimensional quantum-mechanical system, i.e., existing of
 $n$ quantum integrals of motion (operators, which commute with each other as well as with the operator of
 equation~\eqref{1}), simplifies the problem of determining of energy spectrum
and wave functions even when integrability does not lead to the
variables separation, but only to the so-called ``quasi-separation
of variables''~\cite{hudon}.

Therefore, the  classification problem of all potentials~$V$ and
$\vec A$  of electromagnetic field, for which quantum-mechanical
problem is integrable in the sense described above, is of current
importance.

In the three-dimensional case of scalar potential, when there is  no
any magnetic field, this problem was solved as early as in 1967 by
Yakov A. Smorodinsky with coauthors~\cite{smorodinsky,makarov}. They
have proved that if there exist two quantum integrals of motion of
first or second order with respect to $\vec p$ (in the sense
described above) then there exists a possibility of variable
separation in the corresponding Schr\"odinger
equation~\cite{makarov}. The inverse statement is also
true~\cite{smorodinsky}. Thus, eleven classes of integrable
potentials~$V$ obtained by them coincide with the results of classic
paper of Eisenhart~\cite{eisenhart}, where he as early as in 1948
has described all scalar potentials for which corresponding
Schr\"odinger equation (or  Hamilton-Jacobi equation in classical
mechanics) admits variable separation at least in one of eleven
coordinate systems.

The next step was done in 1972 by V.N.~Shapovalov with
coauthors~\cite{shapovalov}. They have obtained complete
classification list of vector-potentials with nonzero magnetic field,
for which corresponding Schr\"odinger equation~\eqref{1} admits
variable separation at least in one of  eleven coordinate systems, and
list of corresponding pairs of operators which commute with each
other and with operator of the equation.

Posterior results in this direction are connected with works of
P.~Winternitz and his coauthors~\cite{dorizzi,mcsween,berube}. For
two-dimensional case they have discovered that existing of
second-order integrals of motion in magnetic field does not
guarantee possibility of variable separation. Nevertheless, even in
this case, integrals of motion are classified on equivalence classes
under action of Euclidian group and second-order terms with respect
to $p_i$ in these integrals have the same form as in the case of
pure scalar potential. It was also shown that in magnetic field
quantum case~\cite{berube} does not obviously coincide with
classical one~\cite{dorizzi,mcsween}, namely, constructed
vector-potentials can depend on Plank constant $\hbar$ in a
nontrivial way.

In this paper we make the next step in classification of potentials
of electromagnetic field~$V$ and $\vec A$ in three-dimensional
Euclidian space, for which corresponding quantum-mechanical system
described by equation~\eqref{1} is integrable in the sense explained
above, i.e., for which there exists a pair of operators~\eqref{2}
that commute with each other as well as with the operator of the
equation. As a result, we obtain a number of vector-potentials for
which corresponding Schr\"odinger equation is integrable but does
not admit variable separation. Therefore, these potentials did not
appear in the Shapovalov's classification~\cite{shapovalov} and are
new.

At first we consider single operator~$Q$ of the form~\eqref{2} which
commutes with the operator  $H$ of Schr\"odinger equation~\eqref{1}.
The commutator $[Q,H]$ contains terms of zero, first, second and third
orders with respect to $p_i$, which coefficients have to vanish.
Coefficients of the third power of $p_i$ give the following system:
\[
\frac{\partial \alpha_{ik}}{\partial x_m}p_mp_ip_k=0.
\]
After solving this system we obtain that the operator~$Q$ can be
presented as symmetric bilinear polynomial of the infinitesimal generators
of a group of motions of three-dimensional Euclidean space $E_3$, i.e.,
symmetry group of the three-dimensional Schr\"odinger equation for free
particle (the Helmholtz equation):
\begin{gather}
Q=  a_{ik}M_iM_k + b_{ik}(p_iM_k+M_kp_i) +c_{ij} p_ip_k+ f_i(x,y,z)
p_i+\gamma_1(x,y,z), \label{3}
\end{gather}
where $a_{ik}$, $b_{ik}$ and $c_{ij}$ are constants, $M_i$ is
operator of rotation, namely, $M_i=\varepsilon_{ikl}x_kp_l$, where
$\varepsilon_{ikl}$ is completely antisymmetric tensor.

Comparing constructed form~\eqref{3} of operator with similar form of operator in the case of
pure scalar potential~\cite{makarov}, we conclude that second-order terms with respect to
 $p_i$ remain the same after appearing of nonzero magnetic field
 (as well as in two-dimensional case~\cite{berube}).

Therefore, analogously to~\cite{makarov}, a pair of commuting operators
 $P$ and $Q$ having the form~\eqref{3} can be reduced by rotations and translations of coordinate system and by
 the transformations
\[
Q'=\mu P +\nu Q +\lambda H
\]
to one of eleven classes  corresponding to eleven classical coordinate systems
which provide separation of variable in the three-dimensional
Schr\"odinger equation for free particle.

That is why in our case of nonzero
magnetic field we obtain that the second-order terms
with respect to $p_i$ in these eleven commuting pairs of the operators  $P$ and
$Q$ are of the same form as ones presented in~\cite{makarov}.
However, in contrast to the pure scalar case, some or all of the coefficients of the
first power of $p_i$ are nonzero functions. The forms of these functions determine the
form of magnetic field (in scalar case they equal zero).

In this paper we completely solve the simplest
``Cartesian'' case, i.e. the case
\begin{gather}
 Q=p_1^2+\vec f(x,y,z)\vec p+\gamma_1(x,y,z),\\[1ex]
P=p_2^2+\vec g(x,y,z)\vec p+\gamma_2(x,y,z).
\end{gather}

Splitting the equations $[H,Q]=[H,P]=[P,Q]=0$ with respect to
different powers of $p_i$, we obtain an overdetermining system of
PDEs on unknown functions $f_1$, $f_2$, $f_3$, $g_1$, $g_2$, $g_3$,
$\gamma_1$, $\gamma_2$ and $V$, $A_1$, $A_2$, $A_3$.

The coefficients of the highest powers of $p_i$ give the following system
\begin{gather*}
 f_2=f_2(x),\quad f_3=f_3(x),\quad g_1=g_1(y),\quad g_3=g_3(y),\quad f_{1y}=g_{2x},\\[1ex]
 f_2'(x)+f_{1y}=4 A_{2x},\quad g_3'(y)+g_{2z}=4 A_{3y},\quad 4 A_{1x}=f_{1x},\\[1ex]
 f_3'(x)+f_{1z}=4 A_{3x},\quad g_1'(y)+g_{2x}=4 A_{1y},\quad 4 A_{2y}=g_{2y}.
\end{gather*}
Its general solution for $\vec A$ is
\begin{gather*}
 4 A_1 = s_x+k_{1x}+g_1(y)+r_1(z),\\[1ex]
 4 A_2 = s_y+k_{2y}+f_2(x)+r_2(z),\\[1ex]
 4 A_3 = s_z+k_{1z}+k_{2z}+f_3(x)+g_3(y)+r_3'(z),
\end{gather*}
where $s=s(x,y,z)$, $k_1=k_1(x,z)$ and $k_2=k_2(y,z)$.

The gauge transformation
\begin{gather*}
 \vec A \rightarrow \vec A + \vec\nabla F,\quad
 F = s(x,y,z) + k_1(x,z) + k_2(y,z) + r_3(z),
\end{gather*}
simplifies the obtained expression for $\vec A$ in the following way
\begin{gather*}
 A_1 = \dfrac 14 (g_1(y)+r_1(z)),\quad
 A_2 = \dfrac 14 (f_2(x)+r_2(z)),\quad
 A_3 = \dfrac 14 (f_3(x)+g_3(y)).
\end{gather*}

With this expression for $\vec A$ in hand, we obtain from the coefficients of the lowest powers of
 $p_i$ the system of ODEs on the functions $g_1(y)$, $r_1(z)$
$f_2(x)$, $r_2(z)$, $f_3(x)$ and $g_3(y)$:
\begin{gather}
 f_2(x)g_3'(y)=g_1(y)f_3'(x),\nonumber\\[1ex]
 r_1(z)f_2'(x)=f_3(x)r_2'(z),\label{4}\\[1ex]
 r_2(z)g_1'(y)=g_3(y)r_1'(z).\nonumber
\end{gather}

It is obvious that the Schr\"odinger equation with
vector-potential~\eqref{1} is invariant with respect to permutations
of $A_1$, $A_2$ and $A_3$, which are done simultaneously with
permutations of the variables
 $x_1$, $x_2$ and $x_3$. Equations~\eqref{4} are invariant with respect to permutations of the functions
 $g_1(y)$,
$r_1(z)$, $f_2(x)$, $r_2(z)$, $f_3(x)$ and $g_3(y)$. These equivalence transformations
can be represented in such a way
\begin{gather*}
\left(\begin{array}{ccc}
f_2 & g_3 & r_1 \\
f_3 & g_1 & r_2 \\
A_1 & A_2 & A_3 \\
x & y & z
\end{array}\right)\quad \sim \quad
\left(\begin{array}{ccc}
r_1 & f_2 & g_3 \\
r_2 & f_3 & g_1 \\
A_3 & A_1 & A_2 \\
z & x & y
\end{array}\right)\quad \sim \quad
\left(\begin{array}{ccc}
g_3 & r_1 & f_2 \\
g_1 & r_2 & f_3 \\
A_2 & A_3 & A_1 \\
y & z & x
\end{array}\right) \quad \sim \quad\\[2ex]
\left(\begin{array}{ccc}
f_3 & r_2 & g_1 \\
f_2 & r_1 & g_3 \\
A_1 & A_3 & A_2 \\
x & z & y
\end{array}\right)\quad \sim \quad
\left(\begin{array}{ccc}
g_1 & f_3 & r_2 \\
g_3 & f_2 & r_1 \\
A_2 & A_1 & A_3 \\
y & x & z
\end{array}\right)\quad \sim \quad
\left(\begin{array}{ccc}
r_2 & g_1 & f_3 \\
r_1 & g_3 & f_2 \\
A_3 & A_2 & A_1 \\
z & y & x
\end{array}\right).
\end{gather*}
Usage of these equivalence transformations allows us to describe exhaustively
all solutions of equations~\eqref{4}. This gives the complete description of all
possible forms of vector-potentials~$\vec A$. The residual coefficients of powers of $p_i$
serve for determining of scalar component~$V$ of vector-potential and put additional constraints on
the functions $g_1(y)$, $r_1(z)$, $f_2(x)$, $r_2(z)$,
$f_3(x)$ and $g_3(y)$.

Below we adduce the final results of our calculations. Here and
below $\vec \Omega$ denotes magnetic field, namely, $ \vec \Omega =
{\rm rot}\, \vec A $.

{\bf Case 1.}
\begin{gather*}
\vec A=0, \quad  \vec \Omega = 0, \quad V  = u_1(x) + u_2(y) + u_3(z),\quad
Q = p_1^2 + 2u_1(x),\quad P = p_2^2 + 2u_2(y).
\end{gather*}
This case corresponds to zero magnetic field and is contained in the
Eisenhart's classification~\cite{eisenhart}. According to his
results all scalar potentials for which corresponding Schr\"odinger
equations admit variable separation in  Cartesian coordinates are
exhausted by ones having the form  \[V = u_1(x) + u_2(y) + u_3(z).\]

{\bf Case 2.}
\begin{gather*}
\vec A= \left(\begin{array}{c}
v_1(z)\\[0.3ex]
v_2(z)\\[0.3ex]
0
\end{array}\right),
\quad \vec \Omega = \left(\begin{array}{c}
-v_2'(z)\\[0.3ex]
\phantom{-}v_1'(z)\\[0.3ex]
0
\end{array}\right),\quad
V  = v_3(z), \quad
Q = p_1^2,\quad
P = p_2^2.
\end{gather*}

 {\bf Case 3.}
\begin{gather*}
\vec A= \left(\begin{array}{c}
0\\[0.3ex]
0\\[0.3ex]
f(x) + g(y)
\end{array}\right),
\quad \vec \Omega = \left(\begin{array}{c}
\phantom{-}g'(y)\\[0.3ex]
-f'(x)\\[0.3ex]
0
\end{array}\right),\quad
V  = u_1(x) + u_2(y),\\[1ex]
Q = p_1^2 + 4 f(x)p_3 + 2u_1(x),\quad
P = p_2^2 + 4 g(y)p_3 + 2u_2(y).
\end{gather*}
The cases 2 and 3 where obtained by Shapovalov et
al.~\cite{shapovalov}. According to their results these two cases
exhaust all vector-potentials with nonzero magnetic field for
which corresponding Schr\"odinger equations admit variable
separation in  Cartesian coordinates.

\medskip
Next three cases are not connected with variable separation and,
therefore, these potentials did not appear in the  Shapovalov's
classification~\cite{shapovalov} and they are new.

{\bf Case 4.}
\begin{gather*}
\vec A= \left(\begin{array}{c}
g'(y)\\[0.3ex]
f'(x)\\[0.3ex]
0
\end{array}\right),
\quad \vec \Omega = \left(\begin{array}{c}
0\\[0.3ex]
0\\[0.3ex]
f''(x)-g''(y)
\end{array}\right),\\[1ex]
 V  = -(C_3f(x)+C_3g(y) +2C_2f^2(x)+2C_1g^2(y)+ r(z)+4g(y)f''(x)+ 4f(x)g''(y)),\\[1ex]
 Q = p_1^2 + 4 f'(x)p_2 - 2(4g(y)f''(x)+2C_2f(x)^2+C_3f(x)),\\[1ex]
P = p_2^2 + 4 g'(y)p_1 - 2(4f(x)g''(y)+2C_1g(y)^2+C_3g(y)),
\end{gather*}
where the functions $f(x)$ and $g(y)$ are solutions of the ODEs
\begin{gather}\label{eq_case4_1}
 f''(x)=Cf^2(x)+C_1f(x)+C_4,\\\label{eq_case4_2}
g''(y)=Cg^2(y)+C_2g(y)+C_5.
\end{gather}
If $C=0$ these are linear second-order ODEs. The case $C\neq0$ is
more interesting. If additionally $C_1\neq0$ (resp. $C_2\neq0$),
then solution of equation~\eqref{eq_case4_1}
(resp.~\eqref{eq_case4_2}) is first Painleve transcendent. If
$C=C_1=0$ (resp. $C=C_2=0$), then~\eqref{eq_case4_1}
(resp.~\eqref{eq_case4_2}) is the Weierstrass equation which
solutions are expressed either via Weierstrass functions or via
elementary ones depending on values of the parameter
 $C_4$ (resp. $C_5$) and integration constants. See, e.g.~\cite{kamke} for details.

{\bf Case 5.}
\begin{gather*}
\vec A= \left(\begin{array}{c}
g'(y)\\[0.3ex]
f'(x)\\[0.3ex]
C f(x)+C g(y)
\end{array}\right),
\quad \vec \Omega = \left(\begin{array}{c}
\phantom{-}C g'(y)\\[0.3ex]
-C f'(x)\\[0.3ex]
f''(x)-g''(y)
\end{array}\right),\\[1ex]
 V  = -(C_3f(x)+C_3g(y) +2C_2f^2(x)+2C_1g^2(y)+ 4g(y)f''(x)+ 4f(x)g''(y)),\\[1ex]
 Q = p_1^2 + 4(f'(x)p_2 +C f(x)p_3) - 2(4g(y)f''(x)+2C_2f(x)^2+C_3f(x)),\\[1ex]
 P = p_2^2 + 4(g'(y)p_1+C g(y)p_3) - 2(4f(x)g''(y)+2C_1g(y)^2+C_3g(y)),
\end{gather*}
where the functions $f(x)$ and $g(y)$ are solutions of the ordinary
differential equations
\begin{gather*}
 f''(x)=C_6f^2(x)+C_1f(x)+C_4,\\
 g''(y)=C_6g^2(y)+C_2g(y)+C_5,
\end{gather*}
which are integrated  analogously to the previous case.

{\bf Case 6.}
\begin{gather*}
\vec A= \dfrac{1}{4}\left(\begin{array}{c}
w_1'(y)+v_1'(z)\\[0.3ex]
u_2'(x)+v_2'(z)\\[0.3ex]
u_3'(x)+w_3'(y)
\end{array}\right),
\quad \vec \Omega = \dfrac{1}{4}\left(\begin{array}{c}
w_3''(y)-v_2''(z)\\[0.3ex]
v_1''(z)-u_3''(x)\\[0.3ex]
u_2''(x)-w_1''(y)
\end{array}\right),\\[1ex]
 V  = -\dfrac 14 (u_1(x)+w_2(y)+v_3(z)+ w_1(y)u_2''(x)+v_1(z)u_3''(x)+{}\\
\phantom{V=}{} u_2(x)w_1''(y)+v_2(z)w_3''(y)+u_3(x)v_1''(z)+w_3(y)v_2''(z)),\\[1ex]
Q = p_1^2 + u_2'(x)p_2 + u_3'(x)p_3 - \dfrac 12 (w_1(y)u_2''(x)+v_1(z)u_3''(x)+u_1(x)),\\[1ex]
 P = p_2^2 + w_1'(y)p_1+ w_3'(y)p_3 -\dfrac 12 (u_2(x)w_1''(y)+v_2(z)w_3''(y)+w_2(y)),
\end{gather*}
where the functions $u_2(x)$, $u_3(x)$, $w_1(y)$, $w_3(y)$,
$v_1(z)$, $v_2(z)$, $u_1(x)$, $w_2(y)$ and $v_3(z)$ are defined in a
special way and described by the following four cases:

{\bf Case 6.1.}
\begin{gather*}
 u_2(x) = a_3(r_1 \cosh(a_1x) + k_1 \sinh(a_1x)),\quad
 u_3(x) = a_2(r_1 \sinh(a_1x) + k_1 \cosh(a_1x)),\\[1ex]
 w_1(y) = a_3(r_2 \cosh(a_2y) + k_2 \sinh(a_2y)),\quad
w_3(y) = a_1(r_2 \sinh(a_2y) + k_2 \cosh(a_2y)),\\[1ex]
v_1(z) = a_2(r_3 \cosh(a_3z) + k_3 \sinh(a_3z)),\quad
 v_2(z) = a_1(r_3 \sinh(a_3z) + k_3 \cosh(a_3z)),\\[1ex]
 u_1(x) = \dfrac {a_2^2a_3^2}4 \left((r_1^2+k_1^2)\cosh(2a_1 x)+ 2r_1k_1\sinh(2a_1x)\right)+
C(r_1 \cosh(a_1 x) + k_1 \sinh(a_1 x)),\\[1ex]
 w_2(y) =
\dfrac {a_1^2a_3^2}4 \left((r_2^2+k_2^2)\cosh(2a_2y)+2r_2k_2\sinh(2a_2y)\right)+
C(r_2 \cosh(a_2y) + k_2 \sinh(a_2y)),\\[1ex]
 v_3(z) =
\dfrac {a_1^2a_2^2}4 \left((r_3^2+k_3^2)\cosh(2a_3z)+2r_3k_3\sinh(2a_3z)\right)+   C_1(r_3 \cosh(a_3z) + k_3
\sinh(a_3z)),
\end{gather*}
with  5 possible subcases:
\begin{alignat*}{5}
&{\rm a)}\quad&&   C=0,\quad&& C_1=0;&& && \\
&{\rm b)}\quad&& r_1 = k_1,\quad&& r_2 = k_2,\quad&& r_3 = k_3,\quad&& C_1=C;\\
&{\rm c)}\quad&& r_1 = k_1,\quad&& r_2 = -k_2,\quad&& r_3 = -k_3,\quad&& C_1=C;\\
&{\rm d)}\quad&& r_1 = -k_1,\quad&& r_2 = k_2,\quad&& r_3 = -k_3,\quad&& C_1=-C;\\
&{\rm e)}\quad&& r_1 = -k_1,\quad&& r_2 = -k_2,\quad&& r_3 = k_3,\quad&& C_1=-C.
\end{alignat*}

{\bf Case 6.2.}
\begin{gather*}
 u_2(x) = a_3(r_1 \sin(a_1 x) - k_1 \cos(a_1 x)),\quad
 u_3(x) = a_2(r_1 \cos(a_1x) + k_1 \sin(a_1x)),\\[1ex]
 w_1(y) = a_3(r_2 \sin(a_2y) - k_2 \cos(a_2y)),\quad
 w_3(y) = a_1(r_2 \cos(a_2y) + k_2 \sin(a_2y)),\\[1ex]
 v_1(z) = a_2(r_3 \cosh(a_3z) + k_3 \sinh(a_3z)),\quad
 v_2(z) = a_1(r_3 \sinh(a_3z) + k_3 \cosh(a_3z)),\\[1ex]
 u_1(x) = \dfrac {a_2^2a_3^2}4 \left((r_1^2-k_1^2)\cos(2a_1x)+2r_1k_1\sin(2a_1x)\right)+
C(r_1 \sin(a_1x) - k_1 \cos(a_1x)),\\[1ex]
w_2(y) =
\dfrac {a_1^2a_3^2}4 \left((r_2^2-k_2^2)\cos(2a_2y)+2r_2k_2\sin(2a_2y)\right)+
C(r_2 \sin(a_2y) - k_2 \cos(a_2y)),\\[1ex]
 v_3(z) =
-\dfrac {a_1^2a_2^2}4 \left((r_3^2+k_3^2)\cosh(2a_3z)+2r_3k_3\sinh(2a_3z)\right)+\\
\phantom{ v_3(z) =\quad} C_1(r_3 \cosh(a_3z) + k_3\sinh(a_3z)),
\end{gather*}
with  5 possible subcases:
\begin{alignat*}{5}
&{\rm a)}\quad&& C=0,\quad&& C_1=0;&& && \\
&{\rm b)}\quad&&  r_1 = i k_1,\quad&& r_2 = -ik_2,\quad&& r_3 = -k_3,\quad&& C_1=i C;\\
&{\rm c)}\quad&& r_1 = i k_1,\quad&& r_2 = i k_2,\quad&& r_3 = k_3,\quad&& C_1=i C;\\
&{\rm d)}\quad&& r_1 = -i k_1,\quad&& r_2 = -i k_2,\quad&& r_3 = k_3,\quad&& C_1=-i C;\\
&{\rm e)}\quad&& r_1 = -i k_1,\quad&& r_2 = i k_2,\quad&& r_3 = -k_3,\quad&& C_1=-i C.
\end{alignat*}

{\bf Case 6.3.}
\begin{gather*}
 u_2(x) = a_3(r_1 \cos(a_1x) + k_1 \sin(a_1x)),\quad
 u_3(x) = ia_2(r_1 \sin(a_1x) - k_1 \cos(a_1x)),\\[1ex]
 w_1(y) = a_3(r_2 \cos(a_2y) + k_2 \sin(a_2y)),\quad
 w_3(y) = ia_1(r_2 \sin(a_2y) - k_2 \cos(a_2y)),\\[1ex]
 v_1(z) = a_2(r_3 \cos(a_3z) + k_3 \sin(a_3z)),\quad
 v_2(z) = ia_1(r_3 \sin(a_3z) - k_3 \cos(a_3z)),\\[1ex]
 u_1(x) = -\dfrac {a_2^2a_3^2}4 \left((r_1^2-k_1^2)\cos(2a_1
x)+2r_1k_1\sin(2a_1x)\right)+
C(r_1 \cos(a_1x) + k_1 \sin(a_1x)),\\[1ex]
 w_2(y) =
-\dfrac {a_1^2a_3^2}4 \left((r_2^2-k_2^2)\cos(2a_2y)+2r_2k_2\sin(2a_2y)\right)+
C(r_2 \cos(a_2y) + k_2 \sin(a_2y)),\\[1ex]
 v_3(z) =
-\dfrac {a_1^2a_2^2}4 \left((r_3^2-k_3^2)\cos(2a_3z)+2r_3k_3\sin(2a_3z)\right)+
C_1(r_3 \sin(a_3z) - k_3\cos(a_3z)),
\end{gather*}
with  5 possible subcases:
\begin{alignat*}{5}
&{\rm a)}\quad&&  C=0,\quad&& C_1=0;&& && \\
&{\rm b)}\quad&& r_1 = i k_1,\quad&& r_2 = -ik_2,\quad&& r_3 = ik_3,\quad&& C_1=i C;\\
&{\rm c)}\quad&& r_1 = -ik_1,\quad&& r_2 = -ik_2,\quad&& r_3 = -ik_3,\quad&& C_1=iC;\\
&{\rm d)}\quad&& r_1 = -ik_1,\quad&& r_2 = ik_2,\quad&& r_3 = ik_3,\quad&& C_1=-i C;\\
&{\rm e)}\quad&& r_1 = ik_1,\quad&& r_2 = ik_2,\quad&& r_3 = -ik_3,\quad&& C_1=-i C.
\end{alignat*}

{\bf Case 6.4.}
\begin{gather*}
 u_2(x) = a_3(r_1 \cosh(a_1x) + k_1 \sinh(a_1x)),\quad
 u_3(x) = -ia_2(r_1 \sinh(a_1x) + k_1 \cosh(a_1x)),\\[1ex]
 w_1(y) = a_3(r_2 \cosh(a_2y) + k_2 \sinh(a_2y)),\quad
 w_3(y) = -ia_1(r_2 \sinh(a_2y) + k_2 \cosh(a_2y)),\\[1ex]
 v_1(z) = a_2(r_3 \cos(a_3z) + k_3 \sin(a_3z)),\quad
 v_2(z) = ia_1(r_3 \sin(a_3z) - k_3 \cos(a_3z)),\\[1ex]
 u_1(x) = \dfrac {a_2^2a_3^2}4 \left((r_1^2+k_1^2)\cosh(2a_1
x)+2r_1k_1\sinh(2a_1x)\right)+
C(r_1 \cosh(a_1 x) + k_1 \sinh(a_1 x)),\\[1ex]
 w_2(y) =
\dfrac {a_1^2a_3^2}4 \left((r_2^2+k_2^2)\cosh(2a_2y)+2r_2k_2\sinh(2a_2y)\right)+
C(r_2 \cosh(a_2y) + k_2 \sinh(a_2y)),\\[1ex]
 v_3(z) =
\dfrac {a_1^2a_2^2}4 \left((r_3^2-k_3^2)\cos(2a_3z)+2r_3k_3\sin(2a_3z)\right)+ C_1(r_3 \sin(a_3z) - k_3
\cos(a_3z)),
\end{gather*}
with  5 possible subcases:
\begin{alignat*}{5}
&{\rm a)}\quad&& C=0,\quad&& C_1=0;&& && \\
&{\rm b)}\quad&& r_1 = -k_1,\quad&& r_2 = -k_2,\quad&& r_3 = -ik_3,\quad&& C_1=C;\\
&{\rm c)}\quad&& r_1 = k_1,\quad&& r_2 = -k_2,\quad&& r_3 = ik_3,\quad&& C_1=C;\\
&{\rm d)}\quad&& r_1 = k_1,\quad&& r_2 = k_2,\quad&& r_3 = -ik_3,\quad&& C_1=-C;\\
&{\rm e)}\quad&& r_1 = -k_1,\quad&& r_2 = k_2,\quad&& r_3 = ik_3,\quad&& C_1=-C.
\end{alignat*}

Therefore, we obtain a number of new vector-potentials, for which the corresponding
Schr\"odinger equation~\eqref{1} is integrable in the sense described above.

\bigskip

The author is grateful to Prof. P. Winternitz for useful discussions.
This work was partly supported by a grant from NATO.


\begin{thebibliography}{99}
\footnotesize\itemsep=-.3ex

\bibitem{hiet1}
Hietarinta J., Pure quantum integrability, {\it Phys. Lett. A} {\bf 246} (1998),
 97--104; arXiv:solv-int/9708010

\bibitem{hudon}  Charest F., Hudon C. and Winternitz P., Quasiseparation of
variables in the Schr\"odinger equa\-tion with a magnetic field,
{\it J. Math. Phys.} {\bf 48} (2007), 012105; arXiv:math-ph/0502046

\bibitem{smorodinsky}
Smorodinskii Ya.A. and Tugov I.I., On complete sets of observables,
{\it J. Exp. Theor. Phys.} {\bf 50} (1966), 653--659 (in Russian);
translated in {\it Soviet Physics JETP} {\bf 23} (1966), 434--436.

\bibitem{makarov}
Makarov A.A., Smorodinsky J.A., Valiev Kh. and Winternitz P., A
systematic search for nonrelativistic systems with dynamical
symmetries, {\it Nuovo Cimento A} {\bf 52} (1967),  1061--1084;
http://dx.doi.org/10.1007/BF02755212

\bibitem{eisenhart}
Eisenhart L.P., Enumeration of potentials for which one--particle
Schr\"odinger equations are separable, {\it Phys. Rev.} {\bf 74} (1948),
 87--89.

\bibitem{shapovalov}
Shapovalov V.N., Bagrov V.G. and Meshkov A.G., Separation of
variables in the stationary Schr\"odinger equation, {\it Izv. Vyssh.
Uchebn. Zaved. Fizika}  {\bf 8} (1972), 45--50 (in Russian);
translated in {\it Russian Physics J.} {\bf 15} (1972), 1115--1119;
http://dx.doi.org/10.1007/BF00910289

\bibitem{dorizzi}
Dorizzi B., Grammaticos B., Ramani A. and Winternitz P., Integrable
Hamiltonian systems with velocity-dependent potentials, {\it J. Math.
Phys.} {\bf 26} (1985),  3070--3079.

\bibitem{mcsween}
McSween E. and Winternitz P., Integrable and superintegrable Hamiltonian
systems in magnetic fields, {\it J.~Math.~Phys.} {\bf 41} (2000),
2957--2967.

\bibitem{berube}
Berube J. and Winternitz P., Integrable and superintegrable quantum
systems in a magnetic field, {\it J. Math. Phys.} {\bf 45} (2004),
1959--1973; arXiv:math-ph/0311051

\bibitem{kamke}
Kamke E., {\it Differentialgleichungen L\"osungsmethoden und
L\"osungen}, Stuttgart, Teubner, 1977.

\end{thebibliography}
\end{document}